\pdfoutput=1
\documentclass[journal]{IEEEtran}

\usepackage{graphicx}
\usepackage{subcaption}
\usepackage{amsmath,amssymb}
\usepackage{booktabs}
\usepackage{xcolor}
\usepackage{url}
\usepackage{cite}
\usepackage{adjustbox}

\begin{document}

\title{
Process-Aware Cross-Layer Adaptation for O-RAN-Enabled Industrial Systems \vspace{-0.1in}
}

\author{Elahe Delavari,~\IEEEmembership{Student Member,~IEEE,}
Junaid Farooq,~\IEEEmembership{Senior Member,~IEEE,}
M. Majid Butt,~\IEEEmembership{Senior Member,~IEEE,}
and Quanyan Zhu,~\IEEEmembership{Senior Member,~IEEE}%
\thanks{E. Delavari and J. Farooq are with the Department of Electrical and Computer Engineering, University of Michigan--Dearborn, Dearborn, MI, USA. Corresponding author: Elahe Delavari (email: elahed@umich.edu).}%
\thanks{M. M. Butt is with Nokia, Naperville, IL 60563 USA (email: majid.butt@nokia.com).}%
\thanks{Q. Zhu is with New York University, Brooklyn, NY, USA (email: qz494@nyu.edu).}
}

\maketitle

\begin{abstract}
Wireless networks increasingly support closed-loop industrial applications in which sensed data must be transmitted, processed, and converted into an action before physical process makes the result obsolete. Throughput, latency, and inference accuracy measured separately cannot determine whether such an application completed a useful task. We propose a Process-Aware Co-adaptation Engine framework that combines application outcomes, process state, radio telemetry, edge-compute state, and sensing configuration to select coordinated operating points across the complete loop. We evaluate the proposed approach in a factory-inspection case study that integrates a physics-based digital twin, a programmable 5G O-RAN network, and edge-based visual inference. The experiments show that the preferred resource allocation changes with production speed and that adapting individual system components independently can be inefficient. We further show that efficient configurations can be identified with relatively few full-system evaluations.
\end{abstract}


\vspace{-0.1in}
\section{Introduction}

\IEEEPARstart {W}{ireless} networks increasingly support closed-loop applications in which sensing, communication, computation, and action must be completed while the underlying physical process continues to evolve. In such systems, successful communication or accurate inference alone does not guarantee a useful outcome; the result must also arrive in time to affect the process. This timing dependence appears in industrial automation, mobile robotics, autonomous vehicles, drone control, remote operation, and extended reality.

Consider, for example, a camera inspecting parts on a moving factory conveyor. The camera captures an image, a wireless uplink carries it to an edge server, an inference service classifies the part, and a robot uses the returned decision to pass or reject it. The classification must be correct, and it must arrive before the part reaches the sorting point. A correct decision that arrives after the part has passed that point has no operational value. We use the term \emph{wireless closed-loop systems} to describe applications in which the usefulness of a result depends on both its correctness and its timing relative to a evolving physical process.

Wireless networked control has long established that communication must be designed with the controlled process in mind~\cite{park2018wireless}. Edge intelligence extends this dependence to computation by placing inference near sensors and making edge capacity part of the end-to-end timing budget~\cite{zhou2019edge}. Task-oriented communications evaluates a link by its contribution to the receiver's task~\cite{gunduz2023beyond}; closed-loop industrial applications extend that task to the physical decision produced by the complete pipeline.
Monitoring remains divided across domains. The radio reports throughput, delay, loss, and channel indicators; the edge reports utilization, queueing, and service latency; the inference pipeline reports accuracy; and the production system reports process state. These signals diagnose local degradation, but none alone establishes that a correct decision reached the actuator while it was still useful.

The physical process also changes the deadline. A faster conveyor shortens the interval between image capture and sorting. A robot moving toward an obstacle reduces the time available for perception and control. Faster user motion in an immersive application reduces the time available to update the displayed scene. The preferred allocation therefore changes even when the wireless channel, edge server, and inference model remain unchanged. A configuration designed for a high process rate may over-provision radio and compute resources when the process slows, whereas one designed for a lower rate may fail to meet the tighter deadline when the process accelerates.
Joint communication-control design addresses part of this coupling~\cite{eisen2019controlaware}, and recent work jointly considers communication, computation, and control in industrial systems~\cite{xia2023cps,diao2025task}. These approaches establish the value of co-design, but a jointly selected design-time configuration does not automatically track changes after deployment. We distinguish \emph{co-design}, which selects cross-layer parameters together for a particular operating condition, from \emph{co-adaptation}, which updates the cross-layer operating point as process and infrastructure states evolve. Co-adaptation becomes necessary when the active bottleneck moves among sensing, radio access, edge computing, and the physical process.

\begin{figure*}[t!]
\centering
\includegraphics[width=0.95\linewidth]{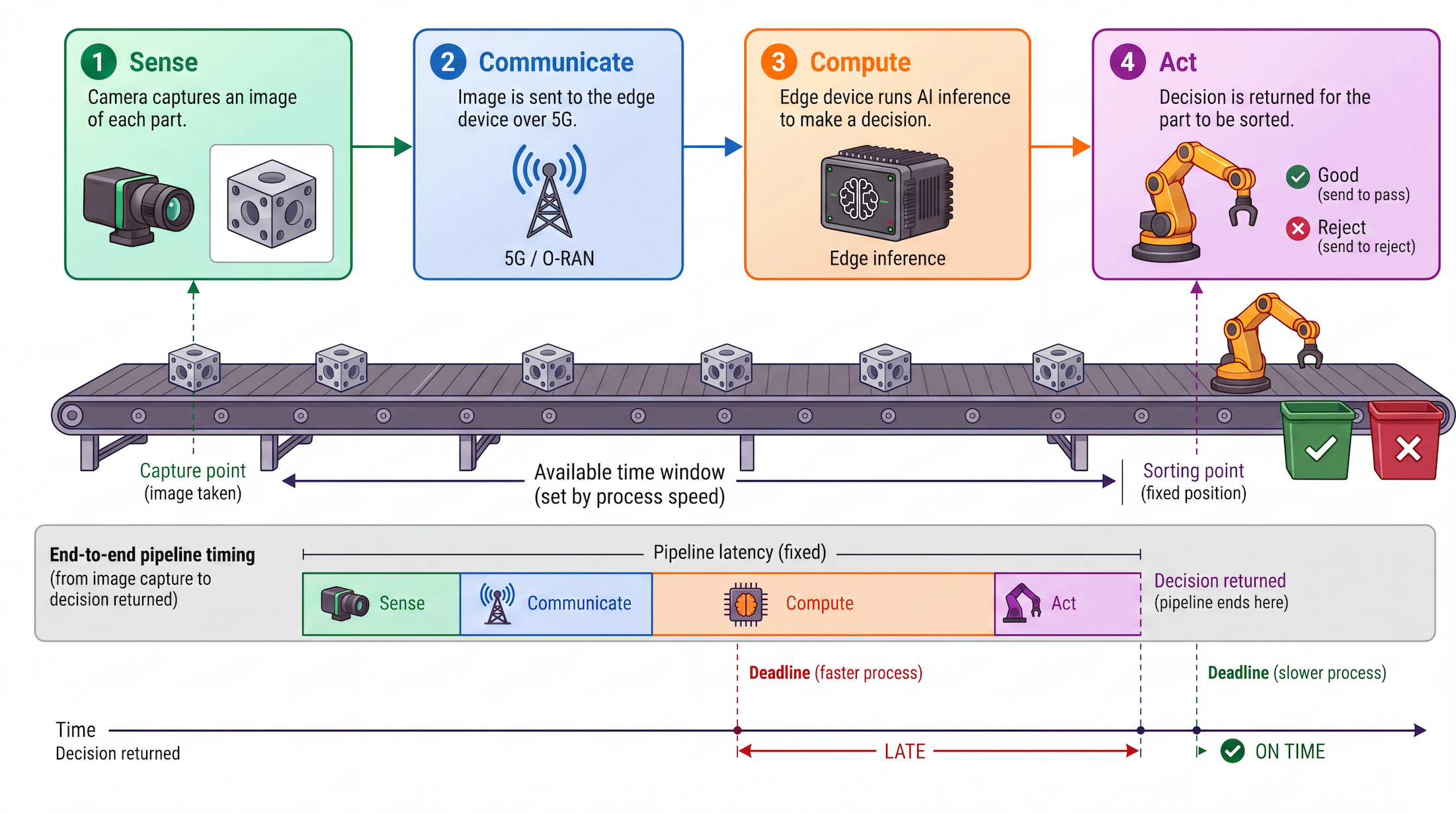}
\caption{\small A process-induced deadline in factory inspection. A part is imaged at the capture point, and the sensing--communication--computation--actuation pipeline must return a decision before the part reaches the fixed sorting point. Increasing conveyor speed shortens the time available to the same end-to-end pipeline.}
\label{fig:idea}
\end{figure*}

Open RAN (O-RAN) provides the radio programmability required for
this transition. Its telemetry and control interfaces expose radio
measurements and enable policy-driven resource adaptation through
RAN intelligent controllers~\cite{polese2023oran}. The application
objective, however, cannot reside exclusively inside the RAN because
the RAN does not observe inference correctness, edge saturation,
sensing configuration, or the process-induced deadline. A
process-aware controller must therefore combine these cross-domain
signals and translate application-level decisions into coordinated
actions across the participating domains. 

We develop a framework, called Process-Aware Co-Adaptation Engine (PACE), that combines radio telemetry, edge-compute state, sensing configuration, process state, and application outcomes to select coordinated cross-layer operating points and translate them into domain-specific control actions. The main contributions of this article are as follows:

\begin{itemize}
    \item We formulate application-level performance metrics for wireless closed-loop systems that jointly account for correctness, timeliness, physical production rate, and shared radio–compute resource use.
    \item We establish a process-aware approach to cross-layer adaptation, in which resource allocation depends on both infrastructure state and the evolving physical process.
    \item We demonstrate these effects in a full-stack factory-inspection system integrating a physics-based digital twin, programmable 5G O-RAN, and edge visual inference.
\end{itemize}

The rest of the article is organized as follows. Section II introduces process-aware performance and application-level metrics. Section III presents the PACE architecture, and Section IV describes the factory-inspection case study. Section V presents the performance evaluation, Section VI discusses open challenges and research directions, and Section VII concludes the article.

\section{Process-Aware Performance}

Process-aware adaptation requires an end-to-end view of application performance. In wireless closed-loop systems, the physical process determines the available timing budget and can shift the limiting stage across sensing, communication, and computation. This section examines these process-induced deadlines and moving bottlenecks, and introduces application-level metrics for evaluating correct, timely, and resource-efficient operation.

\subsection{Deadlines and Moving Bottlenecks}

A conventional communication service is provisioned against externally specified targets for throughput, loss, and delay~\cite{zhang2024tsn}. A closed-loop service instead inherits its timing budget from the physical process. In factory inspection, the decision window equals the capture-to-sort distance divided by conveyor speed, i.e., doubling the speed halves the time available for transfer, inference, decision return, and actuation. Fig.~\ref{fig:idea} illustrates how process rate contracts the shared end-to-end budget.
This budget couples settings managed by different control systems. Image resolution changes visual detail, uplink payload, and inference workload. Uplink physical-resource-block (PRB) allocation influences transfer time, the CPU quota influences inference time, and conveyor speed changes both production output and the deadline. Higher resolution can improve defect visibility while increasing delay; a faster line can raise nominal output while reducing the fraction of actionable decisions.

The binding constraint moves with operating conditions. Under uplink contention, transfer can dominate and additional CPU cores provide little benefit. With high-resolution input on a constrained server, inference can dominate and additional PRBs provide little benefit. Reducing resolution relieves both stages but may reduce classification quality. Effective adaptation must diagnose the current bottleneck before allocating more resources.
Static provisioning reserves capacity for the worst expected condition, protecting deadline margin while consuming resources when the process is slow. Co-design selects the layers jointly for a specific condition, but its operating point becomes stale when production rate, contention, or edge load changes. Co-adaptation preserves the joint view and updates the operating point at runtime.

The process rate can also participate in adaptation: a plant may slow a line when inspection reliability falls or accelerate it when capacity is available. Production and safety controllers must approve such changes because they affect physical throughput, operating targets, and worker safety. Shared infrastructure further favors runtime coordination. Worst-case provisioning for one loop reduces the capacity available to other production lines and industrial applications, whereas co-adaptation seeks the smallest allocation that sustains the required delivered performance.

\begin{figure*}[t]
\centering
\includegraphics[width=0.85\linewidth]{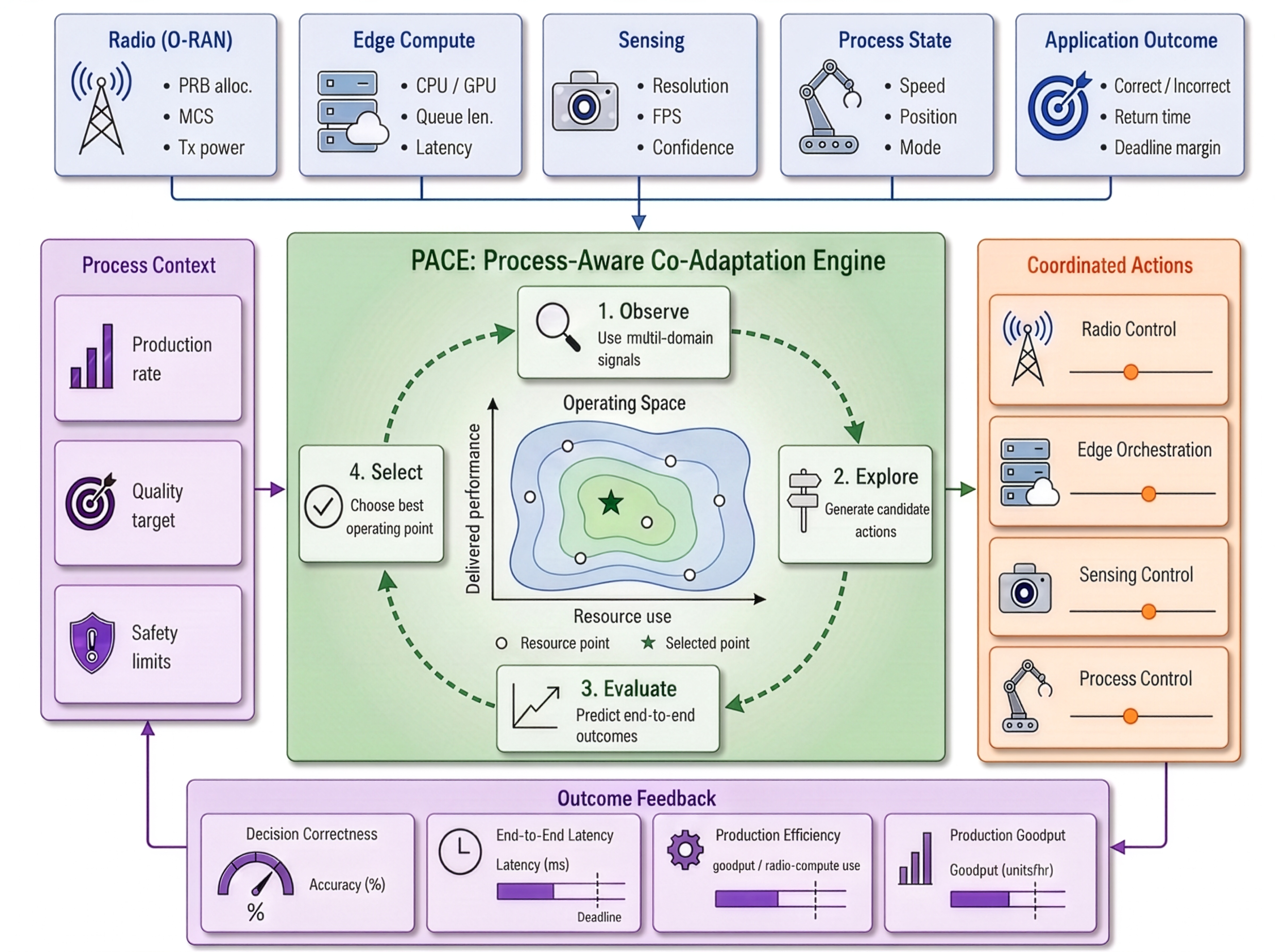}
\caption{\small \textsf{PACE} coordinates sensing and process control, the 5G O-RAN domain, and edge orchestration around an application-level objective. Telemetry and outcomes flow toward the decision engine, while coordinated actions flow to the domain controllers. The near-real-time RIC and E2 interface provide the implemented radio measurement and actuation path.}
\label{fig:architecture}
\end{figure*}

\subsection{Application-Level Metrics}

We evaluate system performance using the outcome of the complete closed loop. High throughput does not guarantee that a decision
beats the process deadline. Low average latency does not
guarantee that the decision is correct. High inference accuracy
does not guarantee that the result remains actionable when it
arrives. Component metrics remain useful for diagnosis, but
the control objective must combine correctness, timeliness,
physical production rate, and shared-resource use.

By combining both computation and communication performance, we define the \emph{correct-on-time rate} as the fraction of inspection cycles in which the returned classification is correct and
available before the part reaches the sorting point. A timely incorrect decision and a correct late decision are both failures. This metric captures the combined effect of sensing quality, wireless delivery, edge processing, and the process-induced deadline, but it does not reward a system for completing more useful cycles per unit time.

We therefore define \textit{production goodput} as the number of parts per minute classified correctly and on time. Correct-on-time rate alone favors slow conveyors because every part receives a generous decision window. Goodput includes the physical production rate, allowing a faster line with a moderately lower success fraction to outperform a slower line that is nearly perfect but produces fewer usable decisions per minute.

Goodput alone can favor configurations that monopolize
the cell and edge server. We define production efficiency, $\eta$ as:
\begin{align}
    \eta = \frac{\text{Production Goodput}}{\text{Uplink Allocation} \times \text{Edge CPU Quota}}.
\end{align}

In the experiments, goodput is expressed in correct parts per minute, uplink allocation is expressed in PRB percentage points, and compute allocation is expressed in CPU cores. The resulting unit is correct parts per minute per
PRB-percentage-per-core. We use this quantity as a relative
efficiency measure within the measured platform; another
deployment can replace the denominator with energy, monetary
cost, GPU time, or a weighted resource price. The multiplicative
denominator penalizes configurations that reserve large radio
and compute allocations simultaneously. 

For each production cycle, the application log records the ground-truth class, returned class, capture and return times, and the deadline determined by the part's motion. These records produce the correct-on-time indicator and aggregate goodput and efficiency. The construction applies beyond manufacturing whenever a deployment can identify the correct outcome, its availability time, and the process state that determines usefulness. Production efficiency does not replace hard application constraints. A controller can enforce a minimum correct-on-time rate before comparing feasible configurations and retain a fallback allocation or reduce process speed when no candidate satisfies the requirement.

\section{\textsf{PACE}: Process-Aware Co-Adaptation Engine}
This section presents the PACE architecture and its process-aware control loop. We describe how PACE combines cross-layer observations and application outcomes to evaluate candidate operating points. We then explain how the selected operating point is translated into coordinated radio, edge, sensing, and process actions operating at different time scales.
\subsection{Cross-Layer Observation and Joint Selection}

We developed a framework, called \textsf{PACE} to close a control loop around delivered application performance. 
PACE controller observes four groups of signals:
\begin{itemize}
    \item \textbf{Radio state:} captures PRB allocation, transfer delay,
    throughput, and contention indicators that characterize the
    communication path.
    \item \textbf{Edge state:} is defined by available compute capacity,
    queueing, and inference latency.
    \item \textbf{Sensing and process state:} comprises image resolution,
    conveyor speed, and the resulting process-induced deadline.
    \item \textbf{Application outcome:} reflects classification correctness,
    return time, deadline satisfaction, and correct-on-time performance,
    indicating whether the complete loop delivered useful work.
\end{itemize}
PACE combines these signals at the application level to construct a cross-layer operating state, as illustrated in Fig.~\ref{fig:architecture}.

The proposed framework converts raw telemetry into a cross-layer operating state. Per-stage timing separates communication delay from inference delay, while process state determines the remaining slack. Application outcomes reveal a confident but incorrect classification or a correct result that missed the sorter, allowing the controller to distinguish radio congestion, compute saturation, sensing limitations, and a tighter process deadline.
PACE then evaluates candidate operating points jointly. Each candidate specifies an uplink PRB share, edge CPU quota, sensing resolution, and an allowed process setting. The selector ranks feasible candidates using correct-on-time production and production efficiency rather than a radio-only or compute-only objective. It rejects candidates that violate application or safety constraints before comparing resource efficiency. This joint decision prevents locally reasonable actions that cannot improve the physical outcome, such as adding PRBs when inference already consumes the remaining deadline.

Selection can use an offline map, a learned model, or both. A measured map provides a direct policy for characterized conditions. Outside that regime, a surrogate model predicts performance and uncertainty across untested configurations, allowing the controller to choose a small number of informative trials rather than sweep the complete space~\cite{garnett2023bo}.
O-RAN provides the implemented radio observation and control path. The near-real-time RIC collects radio measurements and an xApp applies the selected PRB share through the E2 interface~\cite{polese2023oran}.
Longer-horizon deployment policies can enter through the non-real-time RIC and A1, including application priorities, resource budgets, and limits on allowable operating changes. The application controller remains above the RAN because correct-on-time outcomes, edge state, sensing configuration, and process controls lie outside the radio domain.

\subsection{Coordinated Control and Digital-Twin Support}

After selecting an operating point, PACE maps it to domain-specific actions. The radio action changes the industrial traffic's uplink PRB share. The edge action changes the CPU quota or service placement of the inference workload. The sensing action changes image resolution or capture configuration. The process action can request a conveyor-speed change when production policy and safety constraints permit it. Each action retains the semantics and enforcement mechanisms of its own domain, while PACE supplies the common application objective and the coordinated target state.

The actions operate at different time scales. Radio scheduling and allocation can react within the near-real-time RIC window. Edge quotas can change over seconds as workload and queueing evolve. Sensing settings often change after sustained conditions rather than transient fluctuations. Production speed may change only at batch boundaries or after approval from a production controller. PACE therefore uses hierarchical coordination, i.e., fast controllers respond inside constraints established by slower controllers, and slower decisions use aggregated evidence rather than instantaneous fluctuations.


PACE observes telemetry and outcomes, estimates the bottleneck and deadline slack, and selects a feasible cross-layer operating point by evaluating candidate combinations of radio, compute, sensing, and process settings against application and safety constraints. Among the feasible candidates, it favors configurations that sustain correct-on-time performance while using resources efficiently. It then dispatches the corresponding domain actions and verifies the delivered outcome. A failed transition triggers a conservative fallback or a request to reduce the process rate. Within the case-study platform, FlexRIC applies PRB allocations through the near-real-time RIC. We set the edge CPU quota, camera resolution, and conveyor speed through their respective controls before each measurement run. This arrangement evaluates the complete joint operating point over the same end-to-end data path while isolating the effect of each configuration. A production deployment can connect these controls to an orchestration service and production-management interface, subject to the operating limits, approval steps, and safety checks defined by each domain.

The digital twin supports preparation and runtime adaptation~\cite{lin2023dtn}. Offline, it explores combinations that would disrupt a production line and supplies data for the initial operating map and surrogate models. Online, a synchronized twin can rehearse candidate changes. Live feedback remains necessary because channel, traffic, edge, sensing, and physical conditions can drift away from the model. PACE also requires clear control responsibilities and transition rules. The radio controller can change resources within an approved budget, while production and safety controllers review changes to the process rate. Rate limits, dwell times, rollback rules, and conflict checks prevent fast domain controllers from oscillating around slower decisions.

\begin{figure}[t!]
\centering
\includegraphics[width=1\linewidth]{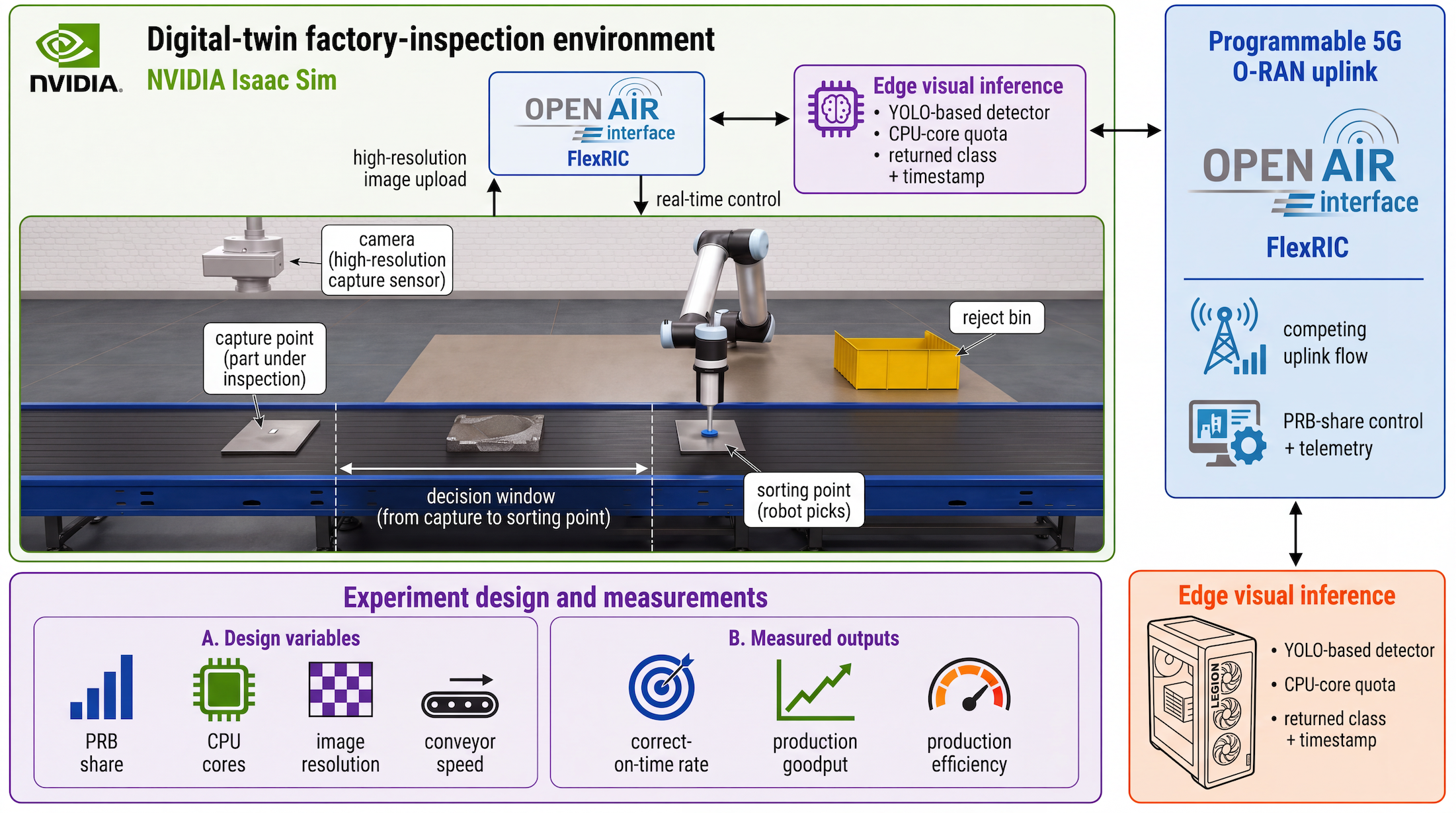}
\caption{\small The digital-twin factory-inspection case study. NVIDIA Isaac Sim generates the moving process and inspection images; an OpenAirInterface 5G NR stack and FlexRIC carry the uplink under contention and control the industrial traffic's PRB allocation; and a fixed YOLO-based edge service returns the defect decision before the part reaches the downstream decision point.}
\label{fig:testbed}
\end{figure}

\section{Case Study: O-RAN-Enabled Factory Inspection Example}

We build a full-stack inspection loop that connects a moving factory process to a contended 5G uplink and edge inference. A part enters the camera's field of view, the industrial user equipment transmits the image, the edge service returns a defect decision, and the application records whether the decision is correct and available before the part reaches the downstream decision point. The integrated path captures protocol behavior, radio contention, inference latency, and the process-induced deadline in the same measurement. Fig.~\ref{fig:testbed} presents the platform and the factory geometry.

\textbf{Physics-based process.} We use Isaac Sim to model the conveyor, part motion, camera, sorting robot, and the geometry that converts conveyor speed into a decision window. The capture point and downstream decision point remain fixed. Increasing conveyor speed therefore shortens the time available without changing the sensing, network, or edge path. Each part triggers a complete measurement cycle beginning at image capture and ending when the classification decision returns.

\textbf{Programmable 5G O-RAN path.} We deploy an OpenAirInterface 5G NR stack with 106 PRBs~\cite{kaltenberger2020oai}. One user equipment carries the inspection images, while a second generates a persistent 30~Mbit/s competing uplink flow. Every configuration is therefore measured under contention rather than on an idle radio link. FlexRIC collects radio telemetry and applies the selected uplink PRB share to the industrial traffic through the near-real-time RIC. We keep the downlink allocation fixed at 50 percent so that the sweep isolates uplink, compute, sensing, and process effects.

\textbf{Edge visual inference.} The edge service crops each received image to the region of interest and processes it with a fixed YOLO-based detector. We train the detector on the public GC10-DET metallic-surface defect dataset~\cite{lv2020deep}. We hold the model architecture and weights fixed throughout the measurements. Differences in delivered performance therefore arise from the selected image resolution, radio allocation, CPU quota, and conveyor speed rather than from changes to the detector.
The design space crosses four axes. The uplink sweep uses 10, 15, 25, 35, 45, 55, 65, 75, 85, and 90 percent of the available PRBs. The edge sweep assigns one through eight CPU cores. The sensing sweep uses $1280{\times}720$, $640{\times}360$, and $320{\times}180$ resolution images. The process sweep uses conveyor speeds of 0.5, 1.0, 1.5, and 2.0~m/s. Their cross-product produces 960 full-system operating points. Each point uses the actual 5G protocol path, the competing uplink, the edge inference service, and the deadline generated by the digital-twin process.

We use conveyor speed in two roles. It is an observed process context when we compare radio, compute, and sensing policies at a requested production rate, and it is a potential control variable when production policy permits PACE to request a rate change. This distinction keeps resource adaptation compatible with production planning: the network and edge first attempt to sustain the requested rate, while process slowdown remains a fallback when no feasible resource configuration satisfies the application constraint.

For every cycle, the application records the ground-truth class, returned classification, timestamps, selected operating point, and deadline outcome. These logs produce the three application metrics and a fixed response map for replaying sequential search policies.
The search replay compares tree-structured Parzen estimation (TPE)~\cite{bergstra2011tpe} and bootstrapped Extra-Trees methods using UCB~\cite{auer2002ucb} and Thompson-style acquisition~\cite{agrawal2012thompson}. These methods generalize across PRB, CPU, resolution, and speed settings.
The latter methods generalize across PRB, CPU, resolution, and speed settings. Policy trajectories follow the feasible transition graph used by the control grid rather than jumping arbitrarily between distant points.

The case study isolates the targeted coupling rather than reproducing every deployment disturbance. The measured deadline ends when the classification decision returns because robotic actuation time was not sufficiently repeatable for inclusion in the sweep. Background traffic and channel conditions remain fixed, and the edge workload does not include time-varying external demand. We also keep the detector weights, capture and decision geometry, competing-flow rate, and downlink allocation unchanged. The numerical optima therefore characterize this platform and context. Fading, mobility, traffic bursts, changing interference, variable compute load, and actuation dynamics require contextual adaptation beyond the fixed operating map.

\section{Performance Evaluation}

\begin{figure}[t]
    \centering
    \includegraphics[width=0.9\linewidth]{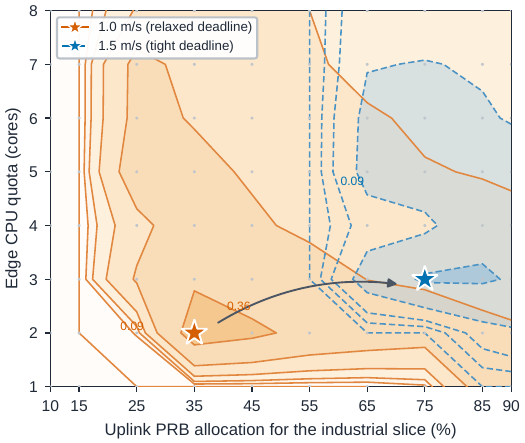}
    \caption{\small Production-efficiency landscape at $1280{\times}720$ resolution as a function of uplink PRB share and edge CPU quota. The stars identify the most efficient measured configurations at conveyor speeds of 1.0 and 1.5~m/s.}
    \label{fig:speed_shift_1280}
\end{figure}

The measured efficiency landscape changes when the conveyor accelerates. As shown in Fig.~\ref{fig:speed_shift_1280}, at a speed of 1.0~m/s, the highest measured production efficiency at $1280{\times}720$ resolution uses a 35 percent uplink share and two CPU cores. At 1.5~m/s, the best measured point moves to a 75 percent uplink share and three cores. The radio environment, edge platform, detector, and image resolution remain unchanged. The shorter process-induced deadline causes the leaner configuration to miss enough decisions that its reduced resource use no longer compensates for the lost goodput.
Provisioning for the faster line wastes resources when the line runs slowly, while provisioning for the slower line fails after acceleration. Process state must therefore enter resource selection; a static mapping from radio KPIs to PRB allocation cannot track the preferred cross-layer point.

\begin{figure}[t]
    \centering
    \adjustbox{trim=0cm 0.2cm 0cm 0cm, clip=true}{%
    \includegraphics[width=\linewidth]{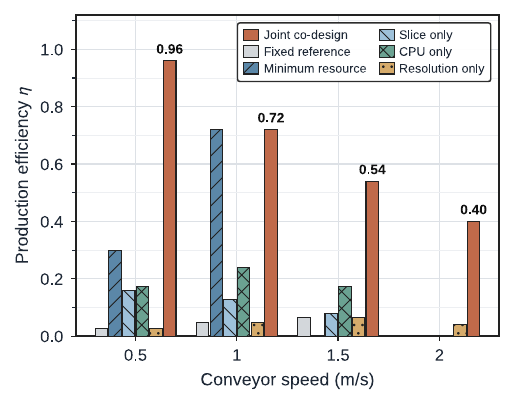}
    }
    \caption{\small Production efficiency of the evaluated allocation policies. Joint selection adapts uplink share, CPU quota, and image resolution together at each conveyor speed; single-parameter policies change one setting while inheriting the remaining settings from the high-resource reference.}
    \label{fig:policy_comparison_efficiency}
\end{figure}

Fig.~\ref{fig:policy_comparison_efficiency} compares joint selection with five reference policies. The fixed high-resource policy uses a 90 percent uplink share, eight CPU cores, and $1280{\times}720$ resolution images at every conveyor speed. The minimum-resource policy uses the lowest radio, compute, and resolution settings. The slice-only policy optimizes the uplink share while retaining the high-resource CPU quota and resolution. The CPU-only policy optimizes compute while retaining the high-resource radio allocation and resolution. The resolution-only policy changes sensing resolution while retaining the high-resource radio and compute allocations. Joint selection searches radio, compute, and resolution together for each process speed.
The fixed high-resource policy achieves an efficiency of approximately 0.027 at 0.5~m/s because it reserves the maximum radio and compute resources for a relaxed deadline. Joint selection reaches 0.96 at the same speed, a gap of about 35 times. The minimum-resource policy reaches 0.72 at 1.0~m/s, where it coincides with the joint optimum, but falls to zero at 1.5~m/s after its deadline margin is exhausted.
Policies adapting only the radio slice, CPU quota, or image resolution never exceed 0.24 in the measured sweep. Joint selection reaches 0.96, 0.72, 0.54, and 0.40 as conveyor speed increases from 0.5 to 2.0~m/s. The decline reflects the tighter deadline and reduced set of configurations that can complete correct decisions on time. The comparison also exposes why isolated adaptation fails. Additional PRBs do not improve the outcome when inference dominates, additional CPU capacity does not improve the outcome when transfer dominates, and a lower resolution may reduce latency while sacrificing the visual information required for correct classification.
A deployed controller cannot test 960 configurations whenever the context changes because each live evaluation consumes production time. 

\begin{figure}[t!]
    \centering
    \adjustbox{trim=0cm 0.2cm 0cm 0cm, clip=true}
    {
    \includegraphics[width=\linewidth]{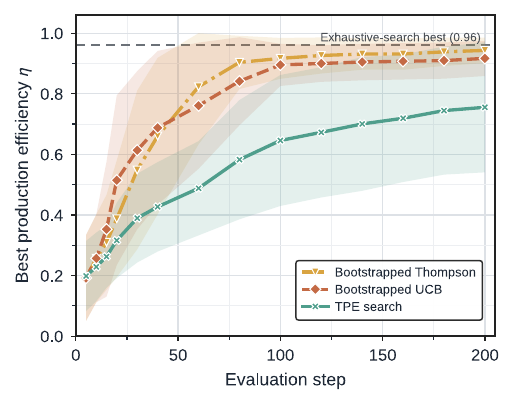}
    }
    \caption{\small Best production efficiency found as the number of transition-constrained full-system evaluations increases. Structure-aware surrogate methods generalize across configurations to identify efficient operating regions with relatively few evaluations. The dotted line marks the best configuration in the measured operating space.}
    \label{fig:eta_search_replay}
\end{figure}

Fig.~\ref{fig:eta_search_replay} shows the replay of sequential policies over the measured map and compare the best efficiency found after each evaluation. The dotted line marks the best efficiency the measured operating space.
Structure-aware methods exploit relationships across the design space. The TPE models promising and unpromising regions, but reaches only about 0.76 after 200 evaluations in this replay. Bootstrapped tree-ensemble UCB and bootstrapped tree-ensemble Thompson-style sampling use ensembles to predict performance and represent uncertainty across neighboring configurations. They reach an efficiency near 0.90 within about 100 evaluations and approach 0.92--0.94 by 200 evaluations, compared with an exhaustive-search best of 0.96. The controller can therefore enter a near-optimal region after evaluating roughly one fifth of the complete 960-configuration design space, and reaches about 90\% of the exhaustive best after roughly one tenth of the grid.



These results outline a practical path toward online co-adaptation. An offline operating map can warm-start a context-aware surrogate, live telemetry can update the estimated operating regime, and constrained acquisition can limit exploration to configurations that satisfy production and safety requirements. When uncertainty increases or feasibility cannot be established, the controller should revert to a verified operating point or reduce the process rate rather than explore blindly. Because the current replay assumes stationary outcomes for each configuration, deployment must additionally detect context shifts, recalibrate uncertainty, and update the surrogate as radio, compute, and process conditions evolve. The central challenge is therefore not merely to find efficient configurations quickly, but to preserve correct-on-time performance while the operating environment changes.

\section{Open Challenges and Research Directions}
This section highlights the key challenges in translating process-aware co-adaptation from a controlled testbed to practical industrial deployment.

\subsection{Coordination Across Control Time Scales}

In wireless closed-loop systems, radio conditions and scheduling can change within milliseconds, edge queues and CPU allocations evolve over seconds, and camera or production settings may remain fixed for minutes or complete batches. A deployable hierarchy should therefore let the near-real-time RIC adjust radio resources within an approved envelope, let the edge orchestrator resize the inference service, and reserve sensing and conveyor changes for the application and production systems. These levels must share a common bottleneck estimate and account for actions already in progress; otherwise, radio and edge controllers may both add resources after observing delay even when a faster conveyor caused the deadline loss. Hysteresis, minimum dwell times, bounded step sizes, and explicit transition states can prevent rapid reversals before the effect of a previous action becomes visible, while stability analysis must account for delayed telemetry and the cycle time needed to observe delivered performance. Coordination policies should also define the permissible range, priority, approval requirement, and rollback condition for each control variable because spectrum, compute, sensing, production, and safety decisions belong to different systems. Each adaptation should record the observed state, requested action, approving system, predicted effect, and delivered outcome so that a local software or telemetry failure cannot silently propagate into the physical process.

\subsection{Safe Adaptation to Changing Conditions}

The measured operating map represents fixed channel conditions, background traffic, edge workload, detector behavior, and process geometry, whereas an operational system will encounter fading, interference changes, traffic bursts, compute contention, camera degradation, and model drift. PACE must determine whether the current conditions remain close enough to the characterized region for reliable interpolation or whether the operating map no longer supports a safe decision, which requires change detection and calibrated uncertainty across radio, compute, sensing, and process variables. Exploration must also reflect the unequal consequences of different controls: changing a PRB share or CPU quota is usually reversible within a short interval, lowering image resolution can reduce inspection quality, and changing conveyor speed can disrupt production planning or safety coordination. A safe policy should therefore expand from a verified operating point, test reversible resource changes first, preserve a known fallback configuration, and reject candidates that violate constraints on correct-on-time rate, resource budgets, or transition magnitude.


When uncertainty is high, the controller should hold the current point, request additional characterization, or ask the production controller to reduce the process rate rather than continue optimizing blindly. A digital twin can reduce live experimentation, but prediction errors in radio contention, inference time, sensing quality, or process dynamics can make an unsafe configuration appear feasible. PACE should continuously compare twin predictions with measured transfer delay, inference latency, correctness, and deadline margin and restrict twin-guided actions when prediction errors exceed calibrated bounds. The central research challenge is to combine offline twin data, sparse live trials, and uncertainty estimates while maintaining a verified safe set as both the environment and its model evolve.

\subsection{End-to-End Monitoring and Evaluation}

Application-level control requires a trustworthy record of every physical cycle. Camera capture, uplink transmission, inference, returned decision, and physical outcome must share a traceable identifier and synchronized timestamps, and the controller must know the age and collection delay of each telemetry stream so that missing or stale observations are not treated as current state. This becomes difficult when measurements cross administrative boundaries or are exposed at different granularities. The resource objective must also remain interpretable, i.e., the current metric isolates PRB and CPU use, but deployments may additionally meter accelerator time, energy, sensing cost, production slowdown, and capacity displaced from other applications. Safety, quality, and deadline requirements should remain hard constraints, with resource cost optimized only over the feasible set and with admission control preventing one closed loop from improving its apparent efficiency by displacing another critical workload. Comparable progress will require benchmarks that preserve this end-to-end context by recording synchronized cycle traces, domain actions, transition costs, resource use, correctness, deadlines, and physical outcomes under changing radio, traffic, compute, sensing, and process conditions. Evaluation should report not only the final operating point but also the deadline misses and production disruption incurred while reaching it. The factory-inspection map provides a controlled foundation, but broader benchmarks must add actuation dynamics and operating changes that the present case study holds fixed.

\section{Conclusion}
We proposed a framework to coordinate sensing, communication, edge computing, and process operation around correct-on-time application performance. The controller combines cross-layer telemetry with delivered outcomes, selects a feasible joint operating point, and translates that point into domain actions operating at different time scales. This application-level loop prevents the system from spending radio or compute resources on a stage that does not determine the current outcome. The factory-inspection case study shows that efficient resource allocation changes with conveyor speed and that joint adaptation outperforms isolated adjustments across the measured operating conditions. Practical deployment requires coordinated cross-domain control, safe adaptation to changing conditions, and end-to-end monitoring of physical outcomes. Addressing these problems would allow O-RAN programmability to become part of a broader process-aware control plane that manages wireless and edge resources according to the useful work they deliver.

\bibliographystyle{IEEEtran}
\bibliography{references}

\end{document}